\documentclass[12pt]{elsarticle}

\usepackage{graphicx}
\usepackage{subfigure}
\usepackage{array}
\usepackage{amssymb}
\usepackage{amsthm}
\usepackage{amsmath}
\usepackage{multirow}
\usepackage{multicol}
\usepackage{lscape}
\usepackage{lineno}
\usepackage{url,color}
\usepackage{bm}
\usepackage{booktabs,colortbl}
\usepackage{resizegather}
\usepackage{adjustbox}
\usepackage{lscape} 
\usepackage{dcolumn}
\usepackage{physics}
\allowdisplaybreaks

\makeatletter
\def\ps@pprintTitle{%
  \let\@oddhead\@empty
  \let\@evenhead\@empty
  \let\@oddfoot\@empty
  \let\@evenfoot\@oddfoot
}
\makeatother

\begin{document}

\begin{frontmatter}
\title{Using Demand Response to Improve Power System Small-Signal Stability}
\author{Mengqi~Yao\corref{cor1}\fnref{label2}}
 \ead{mqyao@berkeley.edu}
 \author{Sandip~Roy\fnref{label3}}
  \ead{sandip@wsu.edu}
\author{Johanna L. Mathieu\fnref{label4}}
\ead{jlmath@umich.edu}
\cortext[cor1]{Corresponding author.}
\fntext[label2]{Mengqi Yao is now a data scientist at Lucid Motors. This work was done when she was a Ph.D. student at the University of Michigan, Department of Electrical Engineering and Computer
Science.}
\fntext[label3]{Washington State University, School of Electrical Engineering and Computer Science}
\fntext[label4]{
University of Michigan, Department of Electrical Engineering and Computer
Science \\
This work was supported by NSF Grants ECCS-1549670 and ECCS-1845093.}


%


\begin{abstract}
With the increase of uncertain and intermittent renewable energy supply on the grid, the power system has become more vulnerable to instability. In this paper, we develop a demand response strategy to improve power system small-signal stability. We pose the problem as an optimization problem wherein the total demand-responsive load is held constant at each time instance but shifted between different buses to improve small-signal stability, which is measured by small-signal stability metrics that are functions of subsets of the system's eigenvalues, such as the smallest damping ratio. To solve the problem, we use iterative linear programming and generalized eigenvalue sensitivities. We demonstrate the approach via a case study that uses the IEEE 14-bus system. Our results show that shifting the load between buses, can improve a small-signal stability margin. We explore the use of models of different fidelity and find that it is important to include models of the automatic voltage regulators and power system stabilizers. In addition, we show that load shifting can achieve similar improvements to generation shifting and better improvement than simply tuning power system stabilizers.  
\end{abstract}


\end{frontmatter}

\section*{Notation}

\noindent\begin{tabular}{ll}
\multicolumn{2}{l}{\hspace{-.2cm}\textbf{Functions}}\\
$f(\cdot)$ &  System dynamic equations\\
$g(\cdot)$ & System algebraic equations \\
$H_{ij}(\cdot)$ & Line flow for line ($i,j$) \\
$h_{ij}(\cdot)$ & Linearization of $H_{ij}$\\
$S_{\mathrm{e},i}(\cdot)$ & Ceiling function of AVR at bus $i$ \vspace{.2cm} \\ 
\multicolumn{2}{l}{\hspace{-.2cm}\textbf{Sets} }\\
\noindent$\mathcal{M}$ & Set of critical eigenvalues\\
\noindent$\mathcal{N}$ & Set of all buses\\
$\mathcal{S}_\mathrm{PV}$ & Set of PV buses\\
$\mathcal{S}_\mathrm{PQ}$ & Set of PQ buses\\
$\mathcal{S}_\mathrm{G}$ & Set of buses with generators\\
$\mathcal{S}_\mathrm{DR}$ & Set of buses with demand-responsive load\\
$\mathcal{S}_\mathrm{PSS}$ & Set of generators with PSS 
\end{tabular}

\noindent \begin{tabular}{p{0.3\columnwidth}p{0.7\columnwidth}}
\multicolumn{2}{l}{\hspace{-.2cm}\textbf{Variables \& Parameters}} \\
$A_{\mathrm{e},i}, B_{\mathrm{e},i}$ &  Ceiling function parameters of AVR at bus $i$ \\
$B_{ij}$ & Susceptance of line $ij$\\
$D_i$ & Damping of gen at bus $i$ \\
$G_{ij}$ & Conductance of line $ij$\\
$ H_i$ & Inertia of gen at bus $i$\\
$K_{\mathrm{a},i}, K_{\mathrm{e},i}, K_{\mathrm{f},i}$ & Parameters of AVR at bus $i$ \\
$K_{\mathrm{w},i}$ & Gain of PSS at bus $i$\\
$i_{\mathrm{d},i}, i_{\mathrm{q},i}$ & dq-axis currents of gen at bus $i$\\
$l_m$ & Left eigenvector of eigenvalue $m$\\
$M$ & Number of finite eigenvalues \\
$p_{\mathrm{d},i}$ & Real power demand at bus $i$\\
$p_{\mathrm{g},i}$ & Real power generation at bus $i$\\
$q_{\mathrm{d},i}$ & Reactive power demand at bus $i$\\
$q_{\mathrm{g},i}$ & Reactive power generation at bus $i$\\
$r_m$ &Right eigenvector of eigenvalue $m$\\
$r_{\mathrm{a},i}$ & Armature resistance of gen at bus $i$\\
$S_n$ & Stability metric $n$\\
$SS_i$ & Numerical Sensitivity with respect to generator $i$\\
$T_{\mathrm{r},i}, T_{\mathrm{a},i}, T_{\mathrm{e},i}, T_{\mathrm{f},i}$ &  Time constants of AVR at bus $i$\\
$T_{\mathrm{w},i},
T_{\mathrm{1},i}, T_{\mathrm{2},i}, T_{\mathrm{3},i}, T_{\mathrm{4},i}$ &  Time constants of PSS at bus $i$\\
$V_i$ & Voltage magnitude at bus $i$\\
$V_{\mathrm{d},i},V_{\mathrm{q},i}$ & dq-axis voltages of gen at bus $i$\\
$V_{\mathrm{f},i}$ & Field voltage of gen at bus $i$\\
$\widetilde{V}_{\mathrm{f},i}$ & Output signal of AVR at bus $i$\\
$V_{\mathrm{m},i}$ & Voltage measured by AVR at bus $i$\\
\end{tabular}

\noindent \begin{tabular}{p{0.3\columnwidth}p{0.7\columnwidth}}
$V_{\mathrm{r1},i},V_{\mathrm{r2},i}$ & Internal signals of AVR at bus $i$\\
$V_{\mathrm{si},i},V_{\mathrm{so},i}$ & Input and output signals of PSS at bus $i$\\
$V_{\mathrm{w},i};V_{\mathrm{p},i}$ & Internal signals of PSS at bus $i$\\
$V_{\textrm{ref},i}$ & Reference terminal voltage of AVR at bus $i$\\
$x'_{\mathrm{d},i}, x'_{\mathrm{q},i} $ & dq-axis transient reactances of gen at bus $i$\\
$x_{\mathrm{w},i}, x_{\mathrm{p},i}, x_{\mathrm{q},i}$ & Internal states of PSS at bus $i$\\
$\bm{x},\bm{x}_{\mathrm{G}}, \bm{x}_{\mathrm{R}},\bm{x}_{\mathrm{S}}$ & Dynamic states\\
$\bm{y}, \bm{y}_{\mathrm{pf}}, \bm{y}_{\mathrm{G}}, \bm{y}_{\mathrm{R}},\bm{y}_{\mathrm{S}}$ &  Algebraic states\\
$\alpha_1$ & Magnitude of largest real part of eigenvalues\\
$\alpha_m$, $\beta_m$ & Real, imaginary parts of eigenvalue $m$\\
$\gamma_n$ & Weighting factor for stability metric $n$\\
$\delta_i$ & Rotor angle of gen at bus $i$\\
$\overline{\varepsilon}, \underline{\varepsilon}$ & Step size maximum, minimum limits\\
$\eta_m$ & Damping ratio of eigenvalue $m$\\
$\eta_\mathrm{S}$ & Smallest damping ratio of generator modes\\
$\eta_\mathrm{I}$ & Damping ratio of critical inter-area mode\\
$\theta_i$ & Voltage angle at bus $i$\\
$\lambda_m$ & Eigenvalue of a matrix\\
$\psi_{\mathrm{d},i},~\psi_{\mathrm{q},i}$ & dq-axis magnetic fluxes of gen at bus $i$\\
$\mu_i$ & Real to reactive power demand ratio at bus $i$\\
$\chi$ & Arbitrary system state\\
$\tau_{\mathrm{m},i}$ & Mechanical torque input of gen at bus $i$ \\
$\omega_i$ & Rotor frequency of gen at bus $i$ \\
\end{tabular}\\

To simplify notation, we assume there is at most one generator (gen) per bus. Superscript `0' denotes setpoint/nominal values. Superscript `*' denotes values at the current operating point. Overlines and underlines denote maximum and minimum limits for the corresponding variable. Bold symbols denote vectors including all variables of a type.

\section{Introduction}
The small-signal  characteristics of the transmission network are strongly influenced by intermittent and uncertain renewable energy sources~\cite{sadamoto2019dynamic}. There is a growing concern that supply fluctuations may cause transmission networks to operate closer to their stability boundaries. Small-signal stability refers to the system's ability to maintain synchronism when subjected to small disturbances \cite{kundur2004definition}. It is well known that tuning power system stabilizers (PSSs) is an effective approach to improve  small-signal stability via increased damping~\cite{jabr2010robust,wang2012novel}. However, tuning PSSs may not be enough to ensure sufficient damping. Generally, PSSs are not tuned in real time because of implementation difficulties. Furthermore, small signal stability issues may be exacerbated in the future with the reduction in power system inertia due to the introduction of higher penetrations of power-electronics-interfaced renewable energy sources. Therefore, multi-faceted strategies to optimize damping in a short period of time are needed.

The operating point of the power network can be shaped to ensure that the network’s small-signal responses
are desirable~\cite{huang2010improving}. This is usually done by generator rescheduling. Some strategies proposed for redispatching generators to improve small-signal stability rely on the Small-Signal Stability Constrained Optimal Power Flow (SSSC-OPF)~\cite{zarate2010opf,chung2004generation,li2013eigenvalue}, which imposes a stability constraint on the optimal power flow solution achieving a lowest-cost operating point that is small-signal stable. Another approach to improve small-signal stability uses sensitivity-based analysis to determine the change in direction of the generation pattern to improve small-signal stability \cite{wang2002increasing,mendoza2013formula,mendoza2016applying}. 

In this paper, we propose a strategy that uses Demand Response (DR) to improve power system small-signal stability. Many demand-responsive loads can respond faster than ramp-limited conventional generators. Fast-acting demand-responsive loads have long been proposed to support power system frequency stability, e.g., \cite{short_stabilization_2007,vrettos2013combined}, and there have also been some proposals to use loads to support voltage stability \cite{yao2019optimal,moors2000design,balanathan1998undervoltage}. Recently there have been a large number of demonstrations of loads providing fast ancillary services \cite{KOHLHEPP2019527,deforest2018day,syed2015ancillary, douglass2011demand}. Some past work has already proposed using DR to improve small-signal stability. For example, \cite{huang2010improving} proposed load shedding to reduce tie-line flows in order to increase the damping of the inter-area oscillation, \cite{tang2017assessment} proposed shifting load in time to improve small-signal stability, and \cite{duan2021hierarchical} proposed an algorithm that co-optimizes generation and demand to enhance small-signal stability while also considering frequency stability. However, changes in system-wide load require corresponding changes in system-wide generation, in order to maintain system frequency, and so the speed of the overall response is limited by the ramp rates of the generators. 

In contrast to load shedding or temporal load shifting, we propose a spatio-temporal load shifting strategy to improve small-signal stability. The strategy reallocates load to different buses while keeping the system-wide loading constant at each time instance, so as not to impact system frequency or require generator redispatch. Spatial load shifting is realized by increasing the energy consumption of flexible loads at some buses while decreasing the energy consumption at others; we do not need to physically shift loads in space. Later in time we will `pay back' the changes.  Over time each load receives as much energy as it would have consumed without participating in DR, but the timing of its consumption is changed (subject to its own flexibility constraints defined by the DR participant). We proposed a similar strategy for improving voltage stability in~\cite{yao2019optimal}. Our strategy could be used when the system does not have an adequate stability margin and generators are unable to respond sufficiently quickly
to correct the problem. In that case, fast-acting demand-responsive loads would respond initially until the slower generators can take over. We anticipate that it would not be cost-effective to develop DR capabilities just for this application, but that loads already capable of fast response for other DR applications could also support small-signal stability, increasing the value to both the power system and DR participant.

The contributions of this paper are as follows. 1) We formulate an optimization problem that uses spatio-temporal loading shifting to optimize small-signal stability metrics that are functions of subsets of the system's eigenvalues.  2) We use iterative linear programming (ILP) and generalized eigenvalue sensitivities \cite{Smed1993,ilea2009damping} to solve the optimization problem. 3) We conduct case studies using the IEEE 14-bus system to determine stability improvements and optimal loading patterns, and we compare the results obtained using different system models -- with and without Automatic Voltage Regulators (AVRs) and PSSs -- to assess the impact and importance of model fidelity.  4)~We compare the performance of spatio-temporal load shifting to generator redispatch, load shedding, and tuning the gain of PSSs to show the relative potential of DR to provide stability-related
services.

This paper substantially builds on our preliminary work \cite{koorehdavoudi2017using} that showed that a small-signal stability margin of Kundur's two-area system \cite{kundur1994power} can be improved by shifting load from area 2 to 1 while keeping the system-wide load constant. However, that paper modeled the load as constant impedance rather than constant power and so stability improved because of spatial load shifting {\em and} the change in load impedance, instead of shifting alone. To focus our investigation on the effectiveness of spatial load shifting, we model the load as constant power in this paper. In addition, that paper used eigenvalue sensitivities corresponding to the reduced system matrix, resulting in high computational cost to solve the problem. Here we use generalized eigenvalue sensitivities, which reduce the computational cost (further details are provided in Section~\ref{Sec: opt}). Furthermore, \cite{koorehdavoudi2017using} did not consider AVRs or PSSs, but since they are commonly used in practice to enhance power system stability \cite{chapman1993stabilizing,dysko2009enhanced}, we consider them here. Under certain circumstances, the AVR/PSS parameters may be incompletely known. Our work investigates whether the inclusion of the models of AVR/PSS will significantly alter the optimal design. Lastly, here we use a larger and more realistic system, i.e., the IEEE 14-bus system, rather than the simpler system used in \cite{koorehdavoudi2017using}.

The remainder of the paper is structured as follows. Section~\ref{Sec: Problem} describes the problem. Section \ref{Sec:model} introduces the models used within the optimization formulation, which is presented in Section \ref{Sec: opt} along with the solution algorithm. Section \ref{Sec: results} details the case studies and Section \ref{Sec: conclusion} gives conclusions.

\section{Problem Description}\label{Sec: Problem}

Poorly damped swing dynamics are a substantial concern in the bulk power grid.  When the system is poorly damped, a small disturbance could lead to the failure of the protection system, the reduction of inter-area power transfer, or more seriously, system instability and cascading failure.  While grid controls and operations are proactively designed to reduce the likelihood of poor-damping scenarios, the power grid is still susceptible to periods of low damping; this susceptibility becomes more pronounced as more intermittent renewables are integrated into the grid, since this increases the variability of system operating points.

Traditionally, damping concerns that arise during operations have been resolved through generation redispatch and, less commonly, other means for shaping the operating solutions (e.g., rerouting of power using new power electronics devices).  However, generation-side power redispatch is slow (taking tens of minutes due to ramping rate constraints), costly, and spatially limited.  In some cases, the bulk power system can degrade quickly toward poor damping and eventually instability, and generator redispatch may have neither the speed nor the spatial support to resolve the damping issues. Hence, alternate methods for resolving damping issues are needed. 

Here, as an alternative, we consider demand-side strategies for resolving poor damping in the bulk grid. Many DR actions can be executed within seconds. Moreover, in many systems, the number of buses with demand responsive loads now exceeds the number of buses with adjustable generators, so demand-side strategies can potentially provide more flexibility to modify the system loading pattern compared with generator redispatch. The following simple example shows the potential benefit of using demand responsive loads to maintain system stability under disturbances.  The example is based on a standard Matlab/Simulink simulation \cite{KundurMatlab} of the Kundur two-area system~\cite{kundur1994power}.

\begin{figure}[t]
\centering
\includegraphics[width=0.8\columnwidth]{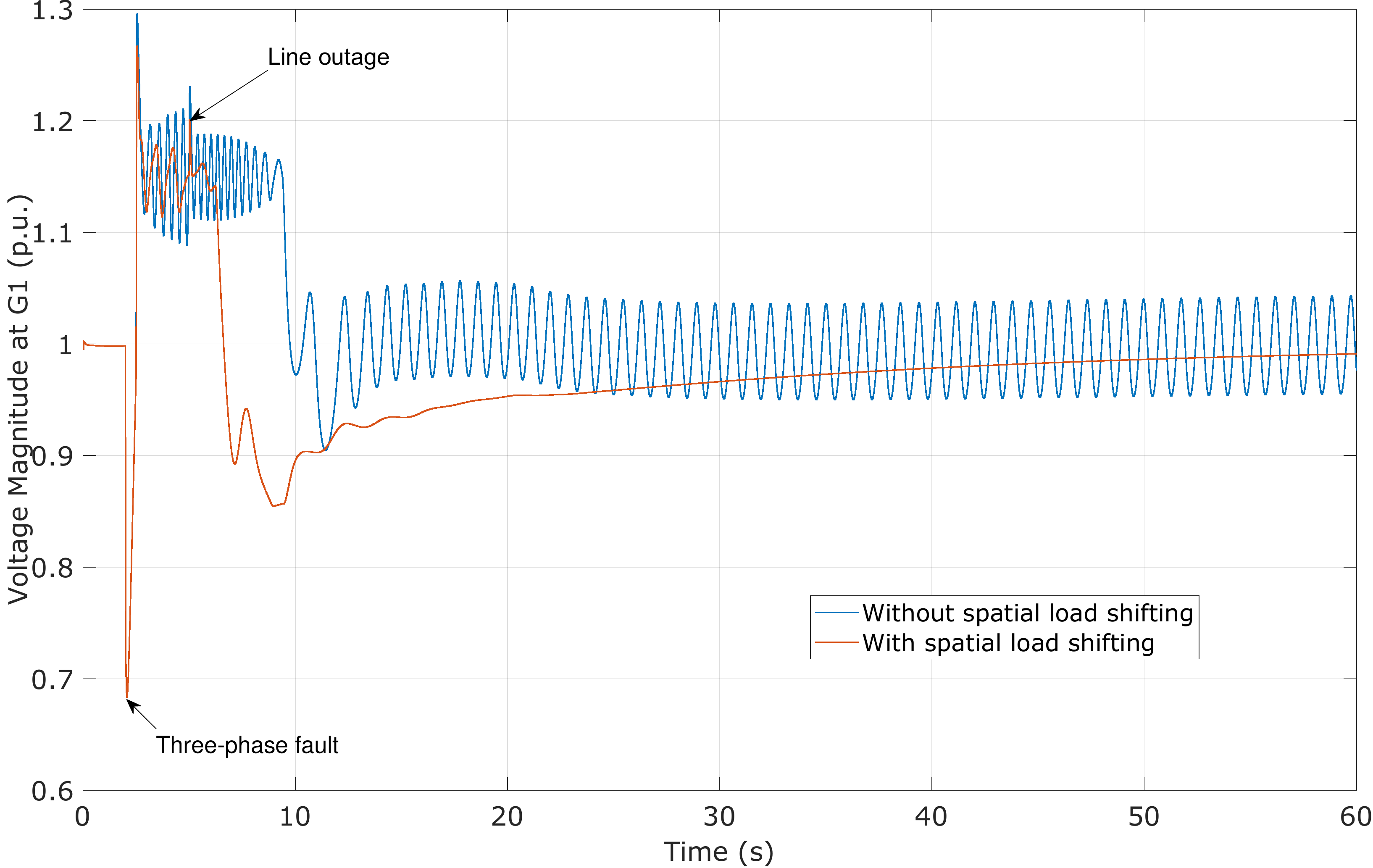}
\vspace{-0.5cm}
\caption{Voltage magnitude at Generator 1 in the Kundur two-area system when a three-phase fault occurs at 2~s and is cleared at 2.5~s, and a line goes out of services at 5~s. Spatial load shifting significantly reduces oscillations.}
\vspace{-0.3cm}
\label{fig:kundur}
\end{figure}

In this example, we focus on the change of the voltage magnitude at Generator~1 after two sequential disturbances.\footnote{A three-phase fault at Bus~7 occurs at 2~s and is cleared at 2.5~s. At 5~s, one of the parallel lines between Bus 7 and Bus 9 goes out of service and stays out of service until the end of the simulation.} This voltage magnitude is shown in blue in Fig.~\ref{fig:kundur}. The system is initially poorly damped and, as a result, the voltage magnitude at Generator~1 has sustained oscillations after the disturbances. The oscillations cannot be eliminated until the disturbances are cleared, or the damping of the system is improved by redispatch. A typical large coal generator may be able to ramp 1\% of its capacity in 1~min~\cite{Vittal2009ImpactOI}; therefore, the generators need minutes to realize the required dispatch to enhance system damping. As an alternative, we have implemented a demand-side strategy for improving the damping.\footnote{The original load is 967 MW in Area 1 and 1767 MW in Area 2. In this simulation, we increase the load in Area 1 by 300 MW and decrease the load in Area 2 by 300 MW, i.e., spatial load shifting between the two areas.} The voltage magnitude at Generator 1 with spatial load shifting is shown in red in Fig.~\ref{fig:kundur}. In this case, since system damping is quickly improved via spatial load shifting, the oscillations are significantly reduced and the system returns to steady state around 1~min after the disturbances.

\begin{figure}[t]
\centering
\vspace{-0.3cm}
\includegraphics[width=0.8\columnwidth]{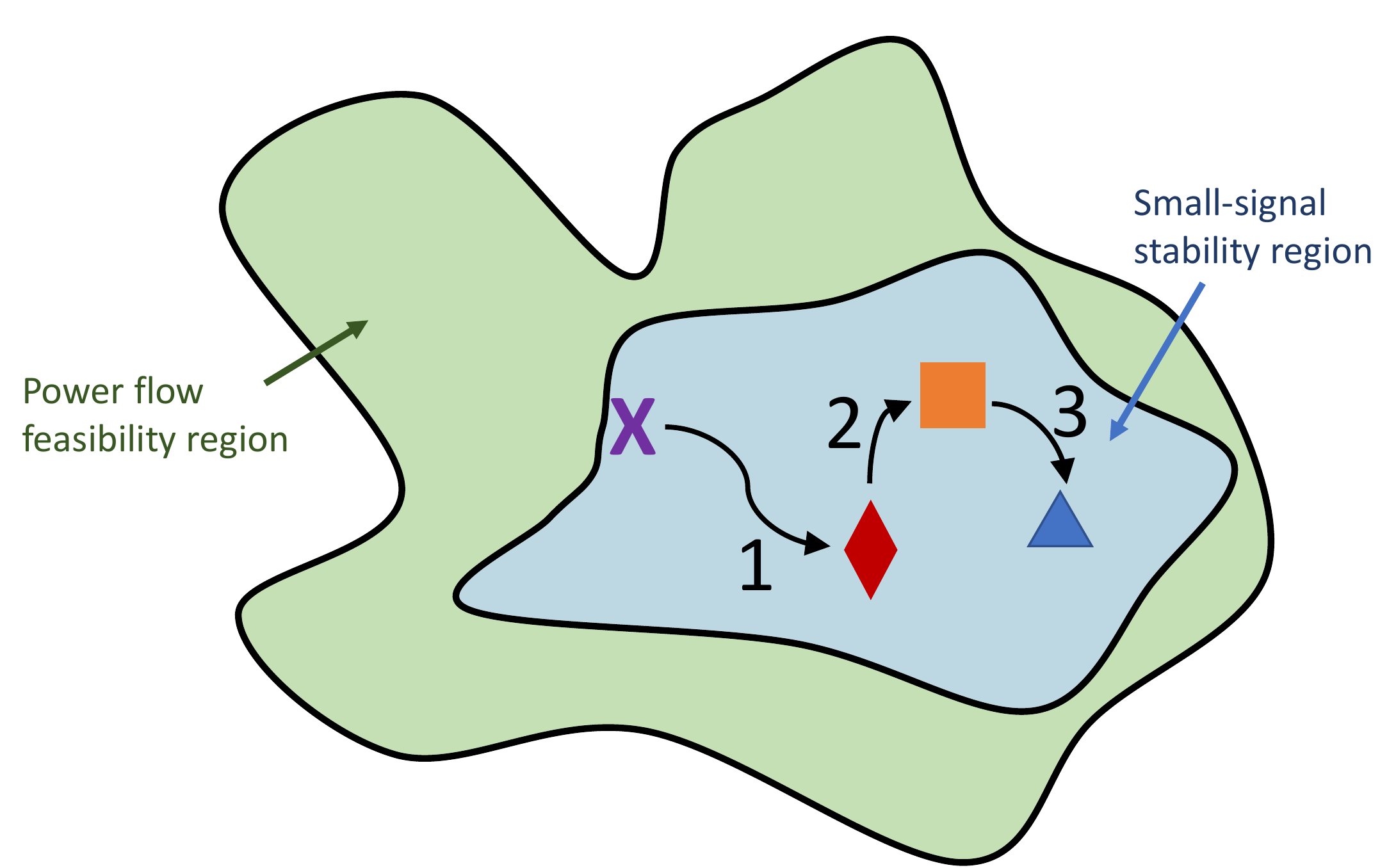}
\caption{Conceptual illustration of the problem. The green and blue areas are the power flow feasibility and small-signal stability regions, respectively. The purple cross is the initial operating point with an inadequate stability margin. The red diamond is the operating point after a demand response action. The orange square is the operating point after the follow-up generator actions. The blue triangle is the operating point after the loads return to their initial values.}
\vspace{-0.3cm}
\label{fig:Problem_des}
\end{figure}

Our aim in this paper is to develop a general methodology for demand-side power shifting to resolve poor damping, based on an optimization formulation.
Figure~\ref{fig:Problem_des} shows a conceptual illustration of the problem. The green shaded region is the feasibility region of power flow (with respect to different generation and load patterns) considering various engineering limits (e.g., line flow, voltage magnitude), and the blue shaded region is the small-signal stability region; they intersect with each other and are both non-convex in general. The initial operating point is shown with the purple cross.  The system is initially operating with an inadequate stability margin. Examples of stability margins include: 1) the difference between the smallest damping ratio (SDR) of the linearized model's eigenvalues and a pre-defined damping ratio threshold for that system \cite{chung2004generation, li2019sequential}; 2) the magnitude of the largest real part among the eigenvalues, i.e., the closest distance of an eigenvalue to the imaginary axis; and 3) the distance of the operating solution to the instability boundary in a suitable state or parameter space. The system operator would like to improve the stability margin by initially only changing the power consumption of fast-acting demand-responsive loads. The system operator shifts load between buses while the total load remains constant. We assume that the active power production of the generators is fixed at the values determined previously via economic dispatch, with the exception of the (fast-acting) generator at the slack bus, which changes its active power production to compensate the change in system losses resulting from the change in loading pattern. Due to the DR action, the operating point moves to the red diamond (Step 1). The loading pattern is maintained for a short period of time while the system operator redispatches slower-acting generation to maintain or improve the new stability margin and also compensate energy to the demand-responsive loads, moving the operating point to the orange square (Step 2). Energy compensation enables loads that reduced (increased) consumption to increase (reduce) consumption for some amount of time to ensure that each load receives as much energy as it would have consumed without participating in DR. Finally, the operating point moves to the blue triangle (Step 3) at which loads return to their initial values and the generators alone ensure small-signal stability.

In this paper, our goal is to determine the optimal dispatch of demand-responsive loads corresponding to the red diamond, i.e., Step~1. Extending our approach to include the other steps would be straightforward, using the same type of  multi-period formulation we proposed for using spatio-temporal load shifting to improve voltage stability in~\cite{yao2019optimal}. However, to focus on the impacts of DR actions on small-signal stability, we only present the formulation and results for Step 1, i.e., single-period spatial load shifting, here.

\section{System Model}\label{Sec:model}
The swing dynamics of a power system can be modeled with a set of nonlinear differential-algebraic equations (DAEs) \cite{Sauerbook}. For simulation and analysis, these equations are often linearized around a current operating point.  Here, we consider stability metrics defined from the linear model -- specifically, functions of the eigenvalues -- which are indicative of the system's damping. Noting that the damping level of the swing dynamics is primarily governed by the inertial responses of the synchronous machines together with control systems including AVRs and PSSs, we use a standard (simplified) model which captures these elements -- the electromechanical dynamics and excitation system, and its associated control systems. In this section, we introduce this model, and in turn give a complete formulation of the design problem, with the aim of highlighting various factors that may influence the spatial load-shifting design.

Formally, we consider a bulk power transmission system with buses belonging to set $\mathcal{N}$.  A subset of the buses have generators and also belong to set $\mathcal{S}_G$. We assume there is at most one generator per bus; this is not a limitation of the approach, but rather for notational simplicity.  One bus in set $\mathcal{S}_G$ is modeled as a slack bus while the others are modeled as PV buses belonging to set $\mathcal{S}_\mathrm{PV}$. The remaining buses in $\mathcal{N}$ are modeled as PQ buses belonging to set $\mathcal{S}_\mathrm{PQ}$. A portion of the buses within $\mathcal{N}$ contain demand responsive loads and those buses belong to set $\mathcal{S}_\mathrm{DR}$.


\subsection{Network and Load Model}

To model the network, we use the AC power flow equations~\cite{Woodbook}
\begin{linenomath}
\begin{subequations}\label{g1}
\begin{align}
     0 =&~V_i\sum_{j \in \mathcal{N}} V_j (G_{ij}\cos(\theta_i-\theta_j)+B_{ij} \sin (\theta_i-\theta_j)) \nonumber\\
      & ~-p_{\mathrm{g},i}+p_{\mathrm{d},i}, \, \forall i \in \mathcal N,\\
      0 =&~V_i\sum_{j \in \mathcal{N}} V_j(G_{ij}\sin(\theta_i-\theta_j)- B_{ij} \cos (\theta_i-\theta_j))\nonumber\\
      & ~ -q_{\mathrm{g},i}+q_{\mathrm{d},i}, \, \forall i \in \mathcal N,  
\end{align}
\end{subequations}
\end{linenomath}
where $V_i$, $\theta_i$ are the voltage magnitude and angle at bus $i$; $G_{ij}$, $B_{ij}$ are the conductance and susceptance of line~$ij$; $p_{\mathrm{g},i}$, $q_{\mathrm{g},i}$ are the real and reactive power generation at bus $i$; and $p_{\mathrm{d},i}$, $q_{\mathrm{d},i}$ are the real and reactive power demand at bus $i$. Here, for simplicity, we assume all loads are constant power loads, though voltage-dependent loads (e.g., represented with a ZIP load model) can easily be included.  For PQ buses,  $p_{\mathrm{g}} = q_{\mathrm{g}} = 0$. Equations~\eqref{g1} are algebraic equations and we define the algebraic state vector associated with power flow as $\bm{y}_{\mathrm{pf},i} = [V_i, \theta_i] \, \forall i \in \mathcal{N}$.

\subsection{Synchronous Machine Model}
A standard differential-algebraic equation model is used for each synchronous machine, which represents its inertial electromechanical dynamics and excitation system, and allows representation of controls.  The differential equations for each machine are~\cite{milano2010power}:
\begin{linenomath}
\begin{subequations}\label{f1}
\begin{align}
   & \dot \delta_i = \omega_i, \, \forall i \in \mathcal S_\mathrm{G}\\
   & \dot \omega_i = \frac{1}{2H_i}\left(\tau_{\mathrm{m},i}-\psi_{\mathrm{d},i} i_{\mathrm{q},i}+\psi_{\mathrm{q},i} i_{\mathrm{d},i} - D_i\omega_i\right), \, \forall i \in \mathcal S_\mathrm{G}
\end{align}
\end{subequations}
\end{linenomath}
where $\delta_i, \omega_i$ are the rotor angle and frequency; $i_{\mathrm{d},i},~i_{\mathrm{q},i}$ are the dq-axis currents; $\psi_{\mathrm{d},i},~\psi_{\mathrm{q},i}$ are the dq-axis magnetic fluxes; and $\tau_{\mathrm{m},i}, D_i, H_i$ are the mechanical torque input, damping, and inertia of the generator at bus $i$. We define the dynamic state vector associated with generators as $\bm{x}_{\mathrm{G},i} = [\delta_i, \omega_i] \, \forall i \in \mathcal{S}_G$,

The algebraic equations for each machine are:
\begin{linenomath}
\begin{subequations}\label{Ch6_g2}
\begin{align}
    & 0 =  V_i\sin(\delta_i-\theta_i) - V_{\mathrm{d},i}, \, \forall i \in \mathcal S_\mathrm{G}, \label{Ch6_g2_st}\\
    & 0 = V_i\cos(\delta_i-\theta_i) - V_{\mathrm{q},i}, \, \forall i \in \mathcal S_\mathrm{G},\label{Ch6_g2_2}\\
   & 0 = V_{\mathrm{d},i} i_{\mathrm{d},i} + V_{\mathrm{q},i} i_{\mathrm{q},i} - p_{\mathrm{g},i} , \, \forall i \in \mathcal S_\mathrm{G}, \label{Ch6_g2_3} \\
   & 0 = V_{\mathrm{q},i} i_{\mathrm{d},i} - V_{\mathrm{d},i} i_{\mathrm{q},i} - q_{\mathrm{g},i} , \, \forall i \in \mathcal S_\mathrm{G}, \label{Ch6_g2_4}\\
    & 0 = \psi_{\mathrm{d},i} +x_{\mathrm{d},i}^{\prime}i_{\mathrm{d},i} - V_{\mathrm{f},i}, \, \forall i \in \mathcal S_\mathrm{G}, \label{elec1}\\
   & 0 = \psi_{\mathrm{q},i} + x_{\mathrm{q},i}^{\prime} i_{\mathrm{q},i}, \, \forall i \in \mathcal S_\mathrm{G}, \label{elec2} \\
    & 0 = -\psi_{\mathrm{d},i} + V_{\mathrm{q},i} + r_{\mathrm{a},i} i_{\mathrm{q},i}, \, \forall i \in \mathcal S_\mathrm{G}, \label{stator1} \\
    & 0 = \psi_{\mathrm{q},i} + V_{\mathrm{d},i} +r_{\mathrm{a},i} i_{\mathrm{d},i}, \, \forall i \in \mathcal S_\mathrm{G},  \label{stator2} \\
   & 0 = V_{\mathrm{f},i} - V_{\mathrm{f},i}^0, \, \forall i \in \mathcal S_\mathrm{G}, \label{G_Vf}
\end{align}
\end{subequations}
\end{linenomath}
where $V_{\mathrm{d},i}, V_{\mathrm{q},i}$ are the dq-axis voltages; $r_{\mathrm{a},i}, V_{\mathrm{f},i}$ are the armature resistance and field voltage; $x_{\mathrm{d},k}^{\prime}, x_{\mathrm{q},k}^{\prime}$ are the dq-axis transient reactances; and $V_{\mathrm{f},i}^0$ is the setpoint of the field voltage of the generator at bus $i$. Equations \eqref{Ch6_g2_st} and \eqref{Ch6_g2_2} link the voltage phasor $V_i \angle \theta_i$ to the dq-axis voltages; \eqref{Ch6_g2_3} and \eqref{Ch6_g2_4} define the real and reactive power injections in terms of the dq-axis voltages and currents;\eqref{elec1} and \eqref{elec2} are the magnetic equations;\eqref{stator1} and \eqref{stator2} are the stator equations; and \eqref{G_Vf} fixes the field voltage to its setpoint. We define the algebraic state vector associated with generators as $\bm{y}_{\mathrm{G},i} = [i_{\mathrm{d},i}, i_{\mathrm{q},i}, V_{\mathrm{d},i}, V_{\mathrm{q},i}, p_{\mathrm{g},i}, q_{\mathrm{g},i}, \psi_{\mathrm{d},i}, \psi_{\mathrm{q},i}, V_{\mathrm{f},i}] \, \forall i \in \mathcal{S}_G$.

Generally, the field voltage $V_{f,i}$ is governed by the AVR. When the AVR model is excluded, this quantity is assumed as fixed, as shown in \eqref{G_Vf}. Otherwise it is governed by the differential equation models for the AVR presented next.

\subsection{Automatic Voltage Regulator (AVR) Model}

\begin{figure}[t]
\centering
  \includegraphics[width=0.8\linewidth]{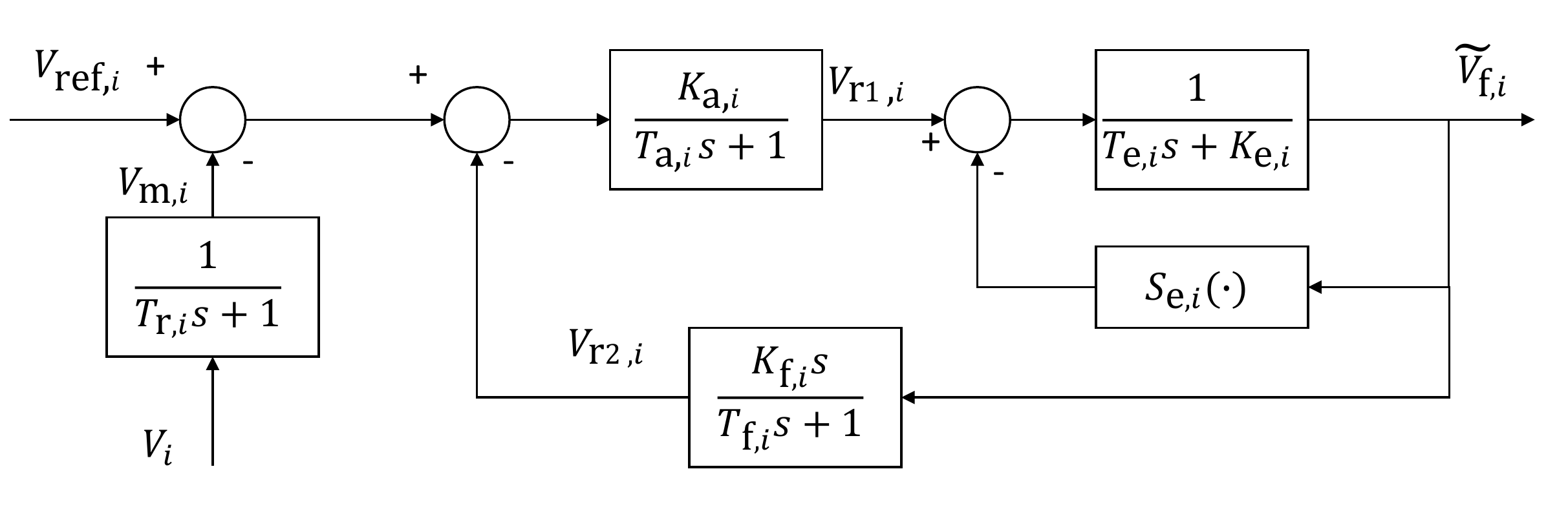}
\caption{AVR block diagram for one generator $i$, based on \cite{milano2010power}.}
\label{fig:AVR}
\end{figure}

AVRs provide primary voltage control \cite{kundur1994power}. We assume each generator has one AVR with the control architecture shown in Fig.~\ref{fig:AVR}. When a generator has an AVR, $V_{\mathrm{f},i}$ in \eqref{G_Vf} is set equal to the output of the AVR; therefore, \eqref{G_Vf} becomes
\begin{linenomath}
\begin{equation}
\label{AVR_extra}
       0 =  \widetilde{V}_{\mathrm{f},i}- V_{\mathrm{f},i}, \, \forall i \in \mathcal S_\mathrm{G}.
\end{equation}
\end{linenomath}

 Based on the block diagram, the differential equations associated with AVRs are~\cite{milano2010power}
 \begin{linenomath}
\begin{subequations}\label{f2}
\begin{align}
   \dot V_{\mathrm{m},i} &= \frac{1}{T_{\mathrm{r},i}}(V_i - V_{\mathrm{m},i}), \, \forall i \in \mathcal S_\mathrm{G},\\
    \dot V_{\mathrm{r1},i} & =\frac{1}{T_{\mathrm{a},i}}(K_{\mathrm{a},i}(V_{\textrm{ref},i}-V_{\mathrm{m},k}-V_{\mathrm{r2},i})-V_{\mathrm{r1},i}), \, \forall i \in \mathcal S_\mathrm{G},\\
    \dot{\widetilde{V}}_{\mathrm{f},i}& = \frac{1}{T_{\mathrm{e},i}}(V_{\mathrm{r1},i}-K_{\mathrm{e},i}\widetilde{V}_{\mathrm{f},i} -S_{\mathrm{e},i}(\widetilde{V}_{\mathrm{f},i})), \, \forall i \in \mathcal S_\mathrm{G},\\
    \dot  V_{\mathrm{r2},i} &= \frac{1}{T_{\mathrm{f},i}}(K_{\mathrm{f},i}\dot{\widetilde{V}}_{\mathrm{f},i}-V_{\mathrm{r2},i}), \, \forall i \in \mathcal S_\mathrm{G},
  \nonumber\\
  & =\frac{1}{T_{\mathrm{f},i}}\left(\frac{K_{\mathrm{f},i}}{T_{\mathrm{e},i}}(V_{\mathrm{r1},i}-K_{\mathrm{e},i}\widetilde{V}_{\mathrm{f},i} -S_\mathrm{e}(\widetilde{V}_{\mathrm{f},i}))-V_{\mathrm{r2},i}\right)
 \end{align}
\end{subequations}
 \end{linenomath}
 where $V_{\mathrm{m},i},V_{\mathrm{r1},i},\widetilde{V}_{\mathrm{f},i},V_{\mathrm{r2},i}$ are the measured voltage, first internal signal, output signal, and second internal signal; $V_{\mathrm{ref},i}$ is the reference terminal voltage;   $T_{\mathrm{r},i}, T_{\mathrm{a},i}, T_{\mathrm{e},i}, T_{\mathrm{f},i}$ are  time constants; and $K_{\mathrm{a},i}$, $K_{\mathrm{e},i}$, $K_{\mathrm{f},i}$ are parameters of the AVR on the generator at bus $i$. The ceiling function  $S_{\mathrm{e},i}(\widetilde{V}_{\mathrm{f},i}) = A_{\mathrm{e},i}e^{B_{\mathrm{e},i}|\widetilde{V}_{\mathrm{f},i}|}$, where $A_{\mathrm{e},i}, B_{\mathrm{e},i}$ are the ceiling function parameters of the AVR on the generator at bus $i$. We define the dynamic state vector associated with AVRs as $\bm{x}_{\mathrm{R},i} = [V_{\mathrm{m},i}, V_{\mathrm{r1},i},  \widetilde{V}_{\mathrm{f},i},V_{\mathrm{r2},i}] \, \forall i \in \mathcal{S}_G$. One more algebraic equation is included,  
\begin{linenomath}
\begin{equation}
 0 = V_{\textrm{ref},i} - V^0_{\textrm{ref},i}, \quad \forall i \in \mathcal S_\mathrm{G}, \label{Vref}
\end{equation}
\end{linenomath}
where $V^0_{\textrm{ref},i}$ is the setpoint of the reference terminal voltage of the AVR on the generator at bus $i$. The reference terminal voltage equation is modified when a PSS is included, as described in the next subsection.

We define the algebraic state vector associated with AVRs as $\bm{y}_{\mathrm{R},i}= [V_{\textrm{ref},i}] \, \forall i \in \mathcal S_\mathrm{G}$.

\subsection{Power System Stabilizer (PSS) Model}
\begin{figure}[t]
\centering
  \includegraphics[width=0.8\linewidth]{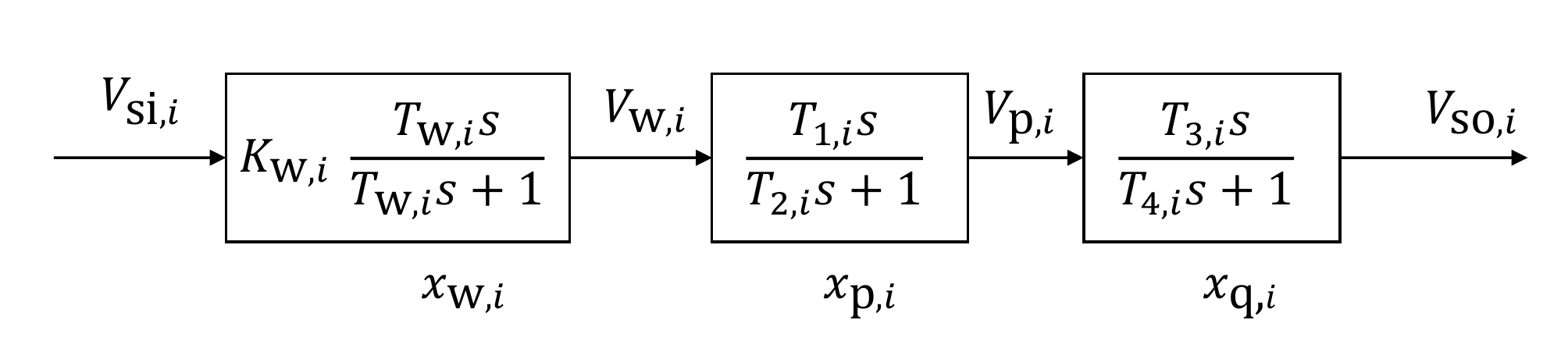}
\caption{PSS block diagram for one generator $i$, based on \cite{kundur1994power}.}
\label{fig:PSS}
\end{figure}

PSSs are used to add damping to generator rotor oscillations \cite{kundur1994power}. We assume a subset of generators have PSSs, where buses with generators with PSSs belong to the set $\mathcal S_\mathrm{PSS}$. 
The output signal $V_{\mathrm{so},i}$ of the PSS modifies the reference terminal voltage $V_{\textrm{ref},i}$ of the AVR and so \eqref{Vref} becomes
\begin{linenomath}
\begin{subequations}\label{PSS_extra}
\begin{align}
   0 &=  V_{\textrm{ref},i} - V^0_{\textrm{ref},i} - V_{\mathrm{so},i}, \, \forall i \in \mathcal S_\mathrm{PSS},\\
    0 &= V_{\textrm{ref},i} - V^0_{\textrm{ref},i}, \, \forall i \in \mathcal S_\mathrm{G} \setminus S_\mathrm{PSS}.
\end{align}
\end{subequations}
\end{linenomath}

Based on the block diagram in Fig.~\ref{fig:PSS}, the differential and algebraic equations associated with PSSs are given as follows \cite{kundur1994power}.  The first block is a high-pass filter and the following two blocks are phase compensators.
\begin{linenomath}
\begin{subequations}\label{f3}
\begin{align}
  & \dot x_{\mathrm{w},i} = \frac{1}{T_{\mathrm{w},i}}V_{\mathrm{w},i}, \, \forall i \in \mathcal S_\mathrm{PSS},\\
   & \dot x_{\mathrm{p},i} = V_{\mathrm{w},i} - V_{\mathrm{p},i}, \, \forall i \in \mathcal S_\mathrm{PSS},\\
   & \dot x_{\mathrm{q},i} = V_{\mathrm{p},i} - V_{\mathrm{so},i}, \, \forall i \in \mathcal S_\mathrm{PSS},
\end{align}
\end{subequations}
\end{linenomath}
where $x_{\mathrm{w},i}, x_{\mathrm{p},i}, x_{\mathrm{q},i}$ are internal states;  $V_{\mathrm{w},i}, V_{\mathrm{p},i}$ are internal signals; $V_{\mathrm{so},i}$ is the output signal, and $T_{\mathrm{w},i}$ is a time constant of the PSS on the generator at bus $i$. We define the dynamic state vector associated with PSS as $\bm{x}_{\mathrm{S},i} = [x_{\mathrm{w},i}, x_{\mathrm{p},i}, x_{\mathrm{q},i}] \, \forall i \in \mathcal{S}_\mathrm{PSS}$.
\begin{linenomath}
\begin{subequations}\label{g4}
\begin{align}
   & 0 = V_{\mathrm{si},i} - K_{\mathrm{w},i} \omega_i, \, \forall i \in \mathcal S_\mathrm{PSS}, \\ 
    &0 = V_{\mathrm{si},i} - V_{\mathrm{w},i}-x_{\mathrm{w},i}, \, \forall i \in \mathcal S_\mathrm{PSS},\\
    & 0 = V_{\mathrm{p},i}T_{2,i}- V_{\mathrm{w},i} T_{1,i}-x_{\mathrm{p},i}, \, \forall i \in \mathcal S_\mathrm{PSS},\\
    & 0 = V_{\mathrm{so},i} T_{4,i} - V_{\mathrm{p},i}T_{3,i} - x_{\mathrm{q},i}, \, \forall i \in \mathcal S_\mathrm{PSS},
\end{align}
\end{subequations}
\end{linenomath}
where $V_{\mathrm{si},i}$ is the input signal; $K_{\mathrm{w},i}$ is the gain; and $T_{1,k}, T_{2,k}, T_{3,k}, T_{4,k}$ are time constants of the PSS on the generator at bus $i$.  We define the algebraic state vector associated with PSSs as $\bm{y}_{\mathrm{S},i} = [V_{\mathrm{si},i}, V_{\mathrm{so},i},$ $V_{\mathrm{w},i}, V_{\mathrm{p},i}] \, \forall i \in \mathcal S_\mathrm{PSS}.$

%

\subsection{Linear State-space Model and Stability Metric}
In general, the full dynamic state vector is  \begin{linenomath}$$\bm{x} =[\bm{x}_{\mathrm{G},i \in \mathcal{S}_\mathrm{G}}, \bm{x}_{\mathrm{R},i \in \mathcal{S}_\mathrm{G}},\bm{x}_{\mathrm{S},i \in \mathcal{S}_\mathrm{PSS}}];$$\end{linenomath} the full algebraic state vector is \begin{linenomath}$$\bm{y} =[\bm{y}_{\mathrm{pf},i \in \mathcal{N}}, \bm{y}_{\mathrm{G},i \in \mathcal{S}_\mathrm{G}}, \bm{y}_{\mathrm{R},i \in \mathcal{S}_\mathrm{G}}, \bm{y}_{\mathrm{S},i \in \mathcal{S}_\mathrm{PSS}}];$$\end{linenomath} function $f$ includes the differential equations \eqref{f1}, \eqref{f2}, and \eqref{f3}; and function $g$ includes the algebraic equations \eqref{g1}, \eqref{Ch6_g2_st}-\eqref{stator2},  \eqref{AVR_extra},\eqref{g4} and \eqref{PSS_extra}. However, we will also explore cases with no PSSs and no AVRs, which removes the corresponding states from these vectors and equations from these functions.

Linearizing the nonlinear DAEs around an operating point yields \cite{kundur1994power}
\begin{linenomath}
\begin{equation}\label{linearmodel}
\begin{bmatrix}
\dot \Delta x\\
0
\end{bmatrix} = 
\underbrace{\begin{bmatrix}
f_x & f_y\\
g_x & g_y
\end{bmatrix}}_A\begin{bmatrix}
\Delta x\\
\Delta y
\end{bmatrix} ,
\end{equation}
\end{linenomath}
where $f_x$, $f_y$, $g_x$, and $g_y$ are the partial derivatives of dynamic and algebraic functions $f,g$ with respect to dynamic and algebraic states $x,y$, respectively.
Define $B = \begin{bmatrix}
I& 0\\
0 & 0
\end{bmatrix}$. Then, \eqref{linearmodel} can be rewritten as
\begin{linenomath}
\begin{equation}\label{systemmatrix}
B \begin{bmatrix}
\dot \Delta x\\
\dot \Delta y
\end{bmatrix} = A \begin{bmatrix}
 \Delta x\\
 \Delta y
\end{bmatrix}.
\end{equation}
\end{linenomath}

We are interested in designing the small-signal characteristics of the power system, which are based on the $M$ finite eigenvalues $\bm \lambda \in \mathbb{R}^M$ of the generalized eigenvalue problem ($A,B$). We index the eigenvalues with $m$, i.e., $\lambda_m$. There are a variety of metrics, defined as functions of subsets of these eigenvalues, used to quantify small-signal stability. Here we list three, though others are possible. Our optimization approach described in the next section could use any of these, or a combination of them.

\subsubsection{Smallest damping ratio (SDR)} The damping of the system is one measure of  disturbance attenuation. Therefore, the smallest damping ratio (SDR) is a common metric for quantifying small-signal stability~\cite{chung2004generation,li2019sequential}. It is defined as
\begin{linenomath}
\begin{equation}
    \eta_\mathrm{S}=\min(\bm \eta),
\end{equation}
\end{linenomath}
where $\bm \eta \in \mathbb{R}^M$ is a vector of damping ratios  $\eta_m=\frac{-\alpha_m}{\sqrt{\alpha_m^2+\beta_m^2}}$, where $\alpha_m=\Re {\lambda_m}$ and $\beta_m=\Im {\lambda_m}$. 

\subsubsection{Damping of the critical inter-area mode}
Another metric is the damping of the critical inter-area mode~\cite{Sauerbook}
\begin{linenomath}
\begin{equation}
    \eta_\mathrm{I}=\frac{-\alpha_I}{\sqrt{\alpha_I^2+\beta_I^2}},
\end{equation}
\end{linenomath}
where  $\alpha_I=\Re {\lambda_I}$ and $\beta_I=\Im {\lambda_I}$, and $\lambda_I$ is the eigenvalue corresponding to one critical mode that has inter-area oscillations.

\subsubsection{Largest real part of the eigenvalues}
A third metric is the magnitude of the largest real part of the eigenvalues~\cite{mallada2013dynamics} 
\begin{linenomath}
\begin{equation}
    \alpha_1 = -\max (\bm \alpha),
\end{equation}
\end{linenomath}
where $\bm \alpha \in \mathbb{R}^M$ is a vector of the real parts of the eigenvalues.

\section{Optimization Approach} \label{Sec: opt}
In this section, we present the formulation of the optimization problem and then introduce the solution algorithm.
\subsection{Formulation}
Our goal is to choose the demand-responsive load $p_{\mathrm{d},i}, q_{\mathrm{d},i}, \, \forall i \in \mathcal S_\mathrm{DR}$ to maximize a linear combination of small-signal stability metrics. Any change to the demand-responsive load will alter the system's operating point, and thus affect the matrix $A$ as well as its eigenvalues and damping ratios. The full optimization problem is
\begin{linenomath}
\begin{subequations}
 \label{nlSDR}
 \begin{align}
  \hspace{-.2cm} \mathrm{max.}  &   \,\,   \sum_n \gamma_n S_n({\bm \lambda} (\bm x, \bm y)) \label{nlSDR_obj}\\
   \mathrm{s.t.}   
     & \,\, g(\bm{x},\bm{y}) = 0 \label{nlSDR_e5}\\
     & \,\, \sum_{i\in \mathcal{S}_\mathrm{DR}} p_{\mathrm{d},i} = \sum_{i\in \mathcal{S}_\mathrm{DR}} p_{\mathrm{d},i}^{0}\label{nlSDR_e6}\\
     &  \,\,   p_{\mathrm{d},i} = \mu_i q_{\mathrm{d},i}   &  \forall i \in \mathcal{N} \label{nlSDR_e7}\\
     & \,\,  p_{\mathrm{d},i} = p_{\mathrm{d},i}^0 & \forall i \in \mathcal{N}\setminus\mathcal{S}_\mathrm{DR}\label{nlSDR_e8}\\
     & \,\, p_{\mathrm{g},i} = p_{\mathrm{g},i}^0 & \forall  i \in \mathcal{S}_\mathrm{PV} \label{nlSDR_e9}\\
     & \,\, V_{i} = V_{i}^0 & \forall  i \in \mathcal{S}_\mathrm{G} \label{nlSDR_e10}\\
     & \,\, \theta_\mathrm{slack} = 0\label{nlSDR_e11}\\
     & \,\, \underline{p}_{\mathrm{d},i} \leq p_{\mathrm{d},i} \leq \overline{p}_{\mathrm{d},i} & \forall i \in \mathcal{S}_\mathrm{DR}\label{nlSDR_nl4}\\
    & \,\, \underline{q}_{\mathrm{d},i} \leq q_{\mathrm{d},i} \leq \overline{q}_{\mathrm{d},i} & \forall i \in \mathcal{S}_\mathrm{DR}\label{nlSDR_nl6}\\
      &  \,\, H_{ij}(\boldsymbol{\theta},\boldsymbol{V}) \leq \overline{{H}}_{ij} & \forall i,j \in \mathcal N \label{nlSDR_niec1}\\
     &\,\, {H}_{ji}(\boldsymbol{\theta},\boldsymbol{V}) \leq  \overline{{H}}_{ji} & \forall j,i \in \mathcal N \label{nlSDR_niec2}\\
& \,\, \underline{p}_{\mathrm{g},\mathrm{slack}} \leq  p_{\mathrm{g},\mathrm{slack}} \leq \overline{p}_{\mathrm{g},\mathrm{slack}}\label{nlSDR_nl1}\\
&\,\,  \underline{q}_{\mathrm{g},i}\leq q_{\mathrm{g},i} \leq \overline{q}_{\mathrm{g},i} & \forall i \in \mathcal{S}_G\label{nlSDR_nl3}\\
&  \,\, \underline{V}_{i} \leq V_{i} \leq \overline{V}_{i} & \forall i \in \mathcal{N} \label{nlSDR_nl5} 
 \end{align}
 \end{subequations}
 \end{linenomath}
where the objective \eqref{nlSDR_obj} is to maximize a weighted linear combination of stability metrics $S_n$ (e.g., $\eta_\mathrm{S}$, $\eta_\mathrm{I}$, and $\alpha_1$), and $\gamma_n$ is the weighting factor corresponding to metric $n$. The stability metrics are functions of the eigenvalues $\bm \lambda$, which in turn are functions of the dynamic and algebraic states $\bm x, \bm y$ defined in the previous section. Constraint \eqref{nlSDR_e5} includes all algebraic equations and will be different for systems with and without AVRs/PSSs. Constraint~\eqref{nlSDR_e6} requires the total demand-responsive load to remain constant, where superscript `0' denotes the nominal value. Constraint \eqref{nlSDR_e7} models all loads as constant power factor loads, where $\mu_i$ is the real to reactive power demand ratio at bus $i$. In Section \ref{Sec: results}, we also explore the performance of spatial load shifting when the real and reactive power consumption of demand-responsive loads are changed independently. Constraints \eqref{nlSDR_e8}-\eqref{nlSDR_e11} fix the real power demand of non-responsive loads, the real power generation at all PV buses, the voltage magnitudes of buses with generators, and the voltage angle of the slack bus to their nominal values. Constraints \eqref{nlSDR_nl4} and \eqref{nlSDR_nl6} limit the flexibility of demand-responsive loads, where underlines and overlines are used to denote lower and upper limits, respectively. Constraints \eqref{nlSDR_niec1}-\eqref{nlSDR_nl5} limit all additional variables that may change with a change in demand-responsive load (i.e., additional decision variables), specifically, line flows, where $H_{ij}$ is a function that computes the flow on line $i,j$ based on the full AC power flow equations; real power generation at the slack bus; reactive power generation at all generators; and voltage magnitudes at all buses. 
\subsection{Solution Algorithm}
The optimization problem \eqref{nlSDR} is challenging to solve in a single shot due to its non-linear, non-convex constraints. Therefore, we use {\em iterative linear programming} (ILP) \cite{Woodbook}. Next, we give an overview of the approach. Then, we describe the method to approximate changes in the eigenvalues as a function of changes in the decision variables and present the linear program solved at each iteration.

The iterative process is described as follows. At each iteration of ILP, we first linearize the objective function and each of the nonlinear constraints at the current operating point $(\bm x^{*},\bm y^{*})$. Then, we solve the resulting linear program (LP) where the new decision variables are the {\em changes} in the original decision variables (e.g., $\Delta p_{\mathrm{d},i},\Delta q_{\mathrm{d},i}, \forall i \in \mathcal S_\mathrm{DR}$). We bound the changes because the linearization is only valid in a small region around the original operating point. Adding the optimal changes to their corresponding values at the current operating point (e.g., $\Delta p_{\mathrm{d},i}+p_{\mathrm{d},i}^*, \Delta q_{\mathrm{d},i}+q_{\mathrm{d},i}^*, \forall i \in \mathcal S_\mathrm{DR}$, where the superscript `*' denotes values at the current operating point) yields an estimate of the solution for the original nonlinear program. However, this estimate may not be a feasible solution to the AC power flow equations. Therefore, the new operating point is computed by solving the AC power flow equations using solution estimate. The process is then repeated, i.e., the nonlinear program is re-linearized around the new operating point to obtain a new linear program, and this linear program is solved to get another estimate of the solution. The iterations continue until the solution estimate converges to the solution (or at least a local maxima) of the original nonlinear program. 

We use generalized eigenvalue sensitivities to approximate the changes in the eigenvalues $\bm \Delta \bm \lambda$ as a function of changes in the decision variables.  For any system state $\chi$ that matrices $A$ and $B$ depend on, the derivative of $\lambda$ with respect to $\chi$ is \cite{Smed1993}
\begin{linenomath}
\begin{equation}\label{generlize_sen}
\frac{\partial \lambda}{\partial \chi} = \frac{l^T(\frac{\partial A}{\partial \chi}-\lambda\frac{\partial B}{\partial \chi})r}{l^TBr},
\end{equation}
 \end{linenomath}
where $r$ and $l$ are the right and left eigenvectors corresponding to $\lambda$. This formula is used to design the HVDC damping controllers in \cite{Smed1993} and FACTS devices in \cite{ilea2009damping}. Since $\frac{\partial B}{\partial \chi} = 0$, we can approximate the change in eigenvalue $\lambda_m$ with respect to a change in all dynamic and algebraic states ($\Delta x_j \forall j, \Delta y_k \forall k$) around  $(\bm x^{*},\bm y^{*})$ as
\begin{linenomath}
\begin{equation}\label{sensitivity}
\Delta \lambda_m = \sum_j \frac{l_m^T\frac{\partial A}{\partial x_j}r_m}{l_m^TBr_m}\Bigg|_{(\bm x^{*}, \bm y^{*})} \hspace{-.8cm} \Delta x_j+\sum_k \frac{l_m^T\frac{\partial A}{\partial y_k}r_m}{l_m^TBr_m}\Bigg|_{(\bm x^{*}, \bm y^{*})} \hspace{-.8cm} \Delta y_k.
\end{equation}
 \end{linenomath}
 
 We next explain why we use generalized eigenvalue sensitivities rather than the more typical approach. Generator modes are usually obtained from the eigenvalues of the reduced matrix $A^{\star} = f_x - f_y(g_y)^{-1}g_x$ \cite{chung2004generation,huang2010improving} rather than the solution of the generalized eigenvalue problem. This is because the reduced matrix has only finite eigenvalues, whereas the generalized eigenvalue problem produces a mixture of finite and infinite eigenvalues. However, to compute eigenvalue sensitivities from $A^{\star}$, one needs to evaluate $\frac{\partial A^{\star}}{\partial \chi}$ numerically (by applying a small disturbance $\Delta \chi$), leading to high computational cost for large systems since large systems mean a large size of $\chi$. In contrast, generalized eigenvalue sensitivities can be expressed analytically, as shown in \eqref{sensitivity}, which is why we use this approach. An alternate approach, which solves the quadratic eigenvalue problem, is used in \cite{mendoza2013formula,mendoza2016applying,mallada2011improving}. The computation complexity is the same as that of the generalized eigenvalue problem; however, the approaches requires the system to be lossless and only allows the reactive power demand to be voltage-dependent. Our approach includes losses and allows both the real and reactive power demand to be voltage-dependent, though here we only investigate cases with constant power loads.

Finally, we can present the LP solved at each iteration. Using the generalized eigenvalue sensitivity \eqref{sensitivity}, we linearize \eqref{nlSDR} about 
 $(\bm x^{*},\bm y^{*})$ resulting in the following optimization problem that must be solved in each iteration.  
 \begin{linenomath}\begin{subequations}\label{linearSDR}
\begin{align}
\hspace{-.3cm} \mathrm{max.}    &  \,\,   \sum_n \gamma_n \Delta S_n(\bm \Delta \bm \lambda (\bm \Delta \bm x, \bm \Delta \bm y))  \label{chap6_ob2}\\
\mathrm{s.t.}    
    & \, \, \eqref{sensitivity} & \forall m \in \mathcal M
    \label{eigensen}\\
    & \,\, \Delta \alpha_m=\Re \{\Delta \lambda_m\},  \hspace{-1.6cm}& \forall m \in \mathcal M \label{deltaalphbet1}\\
    & \,\,  \Delta \beta_m=\Im \{\Delta \lambda_m\}& \forall m \in \mathcal M \label{deltaalphbet2}\\
    &\Delta \eta_m= \frac{(-(\beta_m^*)^2 \Delta \alpha_m+\alpha_m^* \beta_m^* \Delta \beta_m)}{((\alpha_m^*)^2+(\beta_m^*)^2)^{\frac{3}{2}}} \hspace{-1cm}& \forall m \in \mathcal M \label{deltaetaeqn}\\
    & \,\, \sum_j\frac{\partial g}{\partial x_j}\Bigg|_{(\bm x^{*}, \bm y^{*})} \hspace{-.8cm} \Delta x_j  + \sum_k\frac{\partial g}{\partial y_k}\Bigg|_{(\bm x^{*}, \bm y^{*})} \hspace{-.8cm} \Delta y_k = 0 \label{lSDR_e5} \hspace{-6cm}\\
     & \,\, \sum_{i\in \mathcal{S}_\mathrm{DR}} \Delta p_{\mathrm{d},i} = 0\label{lSDR_e6}\\
     &  \,\,   \Delta p_{\mathrm{d},i} = \mu_i \Delta q_{\mathrm{d},i}   &  \forall i \in \mathcal{N} \label{lSDR_e7}\\
     & \,\,  \Delta p_{\mathrm{d},i} = 0 &  \forall i \in \mathcal{N}\setminus\mathcal{S}_\mathrm{DR}\label{lSDR_e8}\\
     & \,\, \Delta p_{\mathrm{g},i} = 0 & \forall  i \in \mathcal{S}_\mathrm{PV} \label{lSDR_e9}\\
     & \,\, \Delta V_{i} = 0& \forall  i \in \mathcal{S}_\mathrm{G} \label{lSDR_e10}\\
     & \,\, \Delta \theta_\mathrm{slack} = 0\label{lSDR_e11}\\
     & \,\, \underline{p}_{\mathrm{d},i} \leq p_{\mathrm{d},i}^* + \Delta  p_{\mathrm{d},i} \leq \overline{p}_{\mathrm{d},i} &\forall i \in \mathcal{S}_\mathrm{DR}\label{lSDR_nl4}\\
     & \,\, \underline{q}_{\mathrm{d},i} \leq q_{\mathrm{d},i}^* + \Delta  q_{\mathrm{d},i} \leq \overline{q}_{\mathrm{d},i} & \forall i \in \mathcal{S}_\mathrm{DR}\\
      &  \,\, {h}_{ij}(\bm x^*, \bm y^*, \boldsymbol{\Delta \theta},\boldsymbol{\Delta V}) \leq \overline{{h}}_{ij} & \forall i,j \in \mathcal N \label{lSDR_niec1}\\
     &\,\, {h}_{ji}(\bm x^*, \bm y^*, \boldsymbol{\Delta \theta},\boldsymbol{\Delta V}) \leq  \overline{{h}}_{ji} & \forall j,i \in \mathcal N\label{lSDR_niec2}\\
& \,\, \underline{p}_{\mathrm{g},\mathrm{slack}} \leq   p_{\mathrm{g},\mathrm{slack}}^* + \Delta p_{\mathrm{g},\mathrm{slack}} \leq \overline{p}_{\mathrm{g},\mathrm{slack}}\label{lSDR_nl1} \hspace{-5cm}\\
&\,\,  \underline{q}_{\mathrm{g},i}\leq q_{\mathrm{g},i}^* + \Delta q_{\mathrm{g},i} \leq \overline{q}_{\mathrm{g},i} & \forall i \in \mathcal{S}_G\label{lSDR_nl3}\\
&  \,\, \underline{V}_{i} \leq V_{i}^* + \Delta V_{i} \leq \overline{V}_{i} &  \forall i \in \mathcal{N} \label{lSDR_nl5} \\
& \,\, \underline{\varepsilon} \leq \Delta \alpha_S \leq \overline{\varepsilon}, \,\, \underline{\varepsilon} \leq \Delta \beta_S \leq \overline{\varepsilon} \hspace{-2cm} \label{lSDR_nl6}
\end{align}
\end{subequations}
\end{linenomath}
where \eqref{eigensen} uses the generalized eigenvalue sensitivity \eqref{sensitivity} to estimate the change in each eigenvalue $\Delta \lambda_m$ in the set of critical eigenvalues $\mathcal M$ (i.e., eigenvalues used to compute stability metrics), ~\eqref{deltaalphbet1} and \eqref{deltaalphbet2} compute the real and imaginary part of $\Delta \lambda_n$, and \eqref{deltaetaeqn} estimates the change in the damping ratios $\Delta \eta_m$. Constraints~\eqref{lSDR_e5} -- \eqref{lSDR_nl5} are the linearization of \eqref{nlSDR_e5} -- \eqref{nlSDR_nl5} and $h_{ij}$ is the linearization of $H_{ij}$. To ensure the accuracy of the linearization, \eqref{lSDR_nl6} is added to limit the size of the change of the eigenvalues, where the step size limits $\underline{\varepsilon}, \overline{\varepsilon}$ can be tuned to improve convergence.

It is possible that when we change the demand-responsive load to improve the damping ratio of the critical eigenvalue (i.e., the eigenvalue with the SDR) it decreases the damping ratio of another eigenvalue so much that that eigenvalue becomes the critical eigenvalue. Therefore, in each iteration we determine which eigenvalues are critical at the current operating point and use these eigenvalues in \eqref{eigensen}-\eqref{deltaetaeqn}. 

The algorithm is terminated when the absolute value of the objective function \eqref{chap6_ob2} goes below a threshold (here, we use $ 10^{-4}$).

\section{Case Studies}\label{Sec: results}
In this section, the IEEE 14-bus system is used to test the effectiveness of our spatial loading shifting strategy on small-signal stability and also illustrate the performance of the ILP approach. We consider only one stability metric, the SDR, since this system does not have inter-area modes. Therefore, the objective function \eqref{nlSDR_obj} is set equal to $\eta_\mathrm{S}$.
The system data and generator parameters can be found in \cite{milano2010power}. The real to reactive power demand ratio is set equal to the nominal ratio, i.e., $\mu_i=p^0_{\mathrm{d},i}/q^0_{\mathrm{d},i} \, \forall i \in \mathcal N$. To clearly demonstrate the impact of the approach, we assume a substantial amount of demand-responsive load. Specifically, we assume the load at all buses except bus~4 (which has the only load with a leading power factor) is demand responsive, resulting in 211.2 MW of demand responsive load out of a total system load of 259 MW. We also assume the load can decrease to 20\% of its nominal value or double, i.e., $\underline{p}_{\mathrm{d},i} = 0.2p_{\mathrm{d},i}$ and $\overline{p}_{\mathrm{d},i} = 2p_{\mathrm{d},i} \, \forall i \in \mathcal S_\mathrm{DR}$. The nominal loading pattern for buses with demand-responsive load is shown in the first row of Table~\ref{tab:14_pd}.  We use step size limits $\underline{\varepsilon}=-0.001, \, \overline{\varepsilon}=0.001$. All computations are implemented in MATLAB on an Intel(R) i5-6600K CPU with 8~GB of RAM. 

\subsection{SDR Improvement \& Optimal Loading Patterns}
\begin{figure}[t]
\centering
\includegraphics[width=5in]{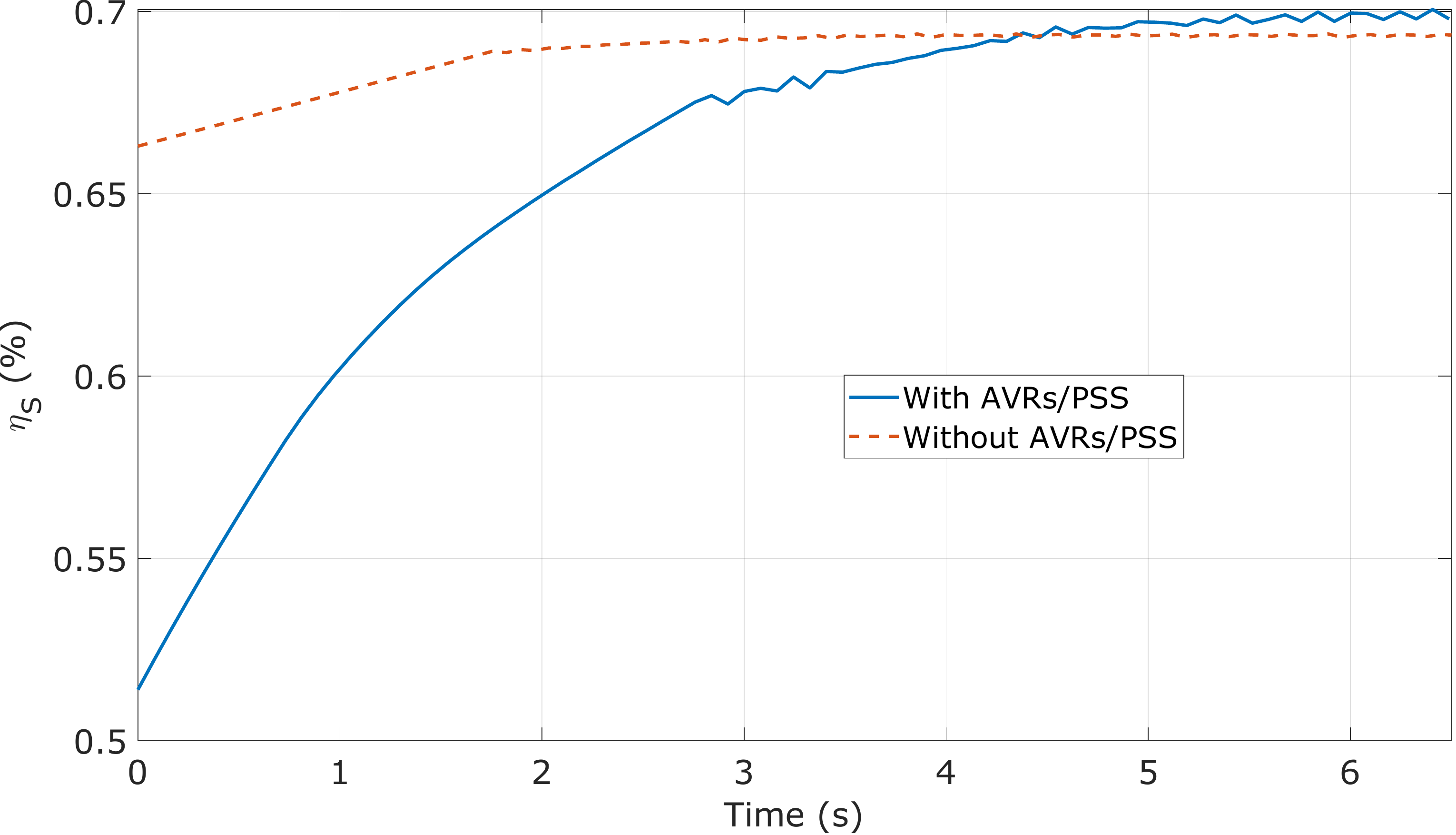}
\vspace{-.8em}
\caption{Convergence of the SDR ($\eta_\mathrm{S}$) for the 14-bus system with AVRs and one PSS (blue solid line), and without AVRs/PSSs (red dashed line).}
\label{fig:ILP_14}
\end{figure}
We first describe the SDR improvement and optimal loading patterns resulting from the use of different system models. Specifically, we consider two models: 1) without AVRs or PSSs, and 2) with AVRs at each generator and a single PSS connected to the generator at bus 1, with parameters $K_{\mathrm{w},1}  = 1$, $T_{1,1} = T_{3,1} = 0.28$, $T_{2,1} = T_{4,1} = 0.02$.\footnote{The AVR and PSS parameters are not optimally tuned, resulting in small damping ratios. However, our results (i.e., qualitative increases in SDRs) are representative and our approach can be applied to general cases.} In addition to showing how the approach improves the SDR by shifting the load, we compare the results corresponding to each model to demonstrate the importance of including AVR/PSS models in the optimization formulation.

We apply the ILP algorithm to each model; Fig.~\ref{fig:ILP_14} shows the convergence of the SDR. The nominal and optimal SDRs are different for the different models. The critical mode without AVRs/PSS corresponds to the generator at bus 2 while with AVRs/PSS it corresponds to the generator at bus 3. The nominal SDR without AVRs/PSS is 0.663\% and the optimal is 0.691\%. The nominal SDR with AVRs/PSS is 0.514\% and the optimal is 0.696\%. We also note that the algorithm converges faster when the system does not have AVRs/PSS. This is not surprising since matrix $A$ is larger with AVRs/PSS and so the computation of the eigenvalue sensitivity \eqref{sensitivity} takes more time.

The optimal loading pattern corresponding to each model is shown in the second and third rows of Table~\ref{tab:14_pd}. Note that the loading patterns are significantly different from each other. If we were to apply the optimal loading pattern determined without considering AVRs/PSS to the system that has AVRs/PSS, the SDR would {\em decrease} from 0.514\% to 0.350\%, implying that accurate system models are very important in determining optimal loading patterns to improve the SDR.

\newcolumntype{d}[1]{D{.}{.}{#1}}

\begin{landscape}
\begin{table}\footnotesize
\centering
\caption{Nominal and optimal loading (MW) at buses with demand-responsive load}
\begin{tabular}{ld{2.2}d{3.2}d{2.2}d{2.2}d{2.2}d{1.2}d{1.2}d{2.2}d{2.2}d{2.2}}
\toprule
  & \multicolumn{1}{c}{Bus 2}     &\multicolumn{1}{c}{Bus 3}       & \multicolumn{1}{c}{Bus 5}      & \multicolumn{1}{c}{Bus 6}      &\multicolumn{1}{c}{Bus 9}      & \multicolumn{1}{c}{Bus 10}   & \multicolumn{1}{c}{Bus 11}    & \multicolumn{1}{c}{Bus 12}     & \multicolumn{1}{c}{Bus 13}     & \multicolumn{1}{c}{Bus 14}     \\
  \midrule
Nominal, $p_\mathrm{d}^0$, from \cite{milano2010power} & 21.70 & 94.20  & 7.60  & 11.20 & 29.50 & 9.00 & 3.50 & 6.10  & 13.50 & 14.90 \\
Optimal, without AVRs/PSS & 4.34  & 141.17 & 5.92 & 13.18  & 5.90  & 1.80 & 0.70 & 30.65 & 4.56  & 2.98  \\
Optimal, with AVRs/PSS & 26.38 & 58.09  & 66.37 & 26.58 & 12.1  & 4.82 & 3.24 & 3.04  & 4.08 & 6.51 \\
Load shedding, with AVRs/PSS & 20.95 & 75.18  & 7.33 & 10.83 & 28.48 & 8.69 & 3.35 & 5.66  & 12.52  & 13.82\\
Optimal, with AVRs/PSS and PSS tuning & 22.06 & 48.55  & 84.98 & 24.36 & 13.72 & 9.53 & 0.77& 1.39  & 2.75  & 3.09\\
\bottomrule
\end{tabular}
\label{tab:14_pd}
\end{table}
\end{landscape}
\begin{table}[t]\footnotesize
\caption{Performance comparison across cases with different decision variables and/or demand-responsive load constraints}
\centering
\vspace{1em}
\begin{tabular}{l  c c c c c c c }
\toprule
Case & 1 & 2 & 3 & 4 & 5 & 6 & 7  \\
\midrule
{\bf Decision Variables}\\
$p_{\mathrm{g}}$ & & & & & &$\checkmark$  & $\checkmark$  \\
$p_{\mathrm{d}}$ &$\checkmark$ &$\checkmark$ & &$\checkmark$&$\checkmark$ &  &$\checkmark$  \\
$q_{\mathrm{d}}$ & $\checkmark$ &  & $\checkmark$&$\checkmark$&$\checkmark$ &  &$\checkmark$ \\
\midrule
{\bf DR Load Constraints}\\
$p_\mathrm{d}=\mu q_\mathrm{d}$ & $\checkmark$ &  & & & &    &\\
$|q^0_{\mathrm{d}}- q_{\mathrm{d}}| \leq 100$MVar &  &  & $\checkmark$&$\checkmark$ & &    &\\
$|q^0_{\mathrm{d}}-q_{\mathrm{d}}| \leq 20$MVar &  &  & & &$\checkmark$& &  $\checkmark$ \\
\midrule
{\bf Results}\\
Optimal SDR (\%) & 0.696 & 0.702 & 0.638 & 0.720 & 0.704 &0.726  & 0.768\\
 Percent Improvement (\%) &35.4 & 36.5 & 24.1 & 40.1 &36.9 & 41.2& 49.5\\
\bottomrule
\end{tabular}
\label{table:sdr_dv}
\end{table}

\subsection{Comparison of SDR Improvement}
We next compare seven cases with different decision variables and/or demand-responsive load constraints; case descriptions and results are shown in Table~\ref{table:sdr_dv}. The results shed light on the relative value of shifting real versus reactive power demand and demand versus generation, and the impact of the demand-responsive load characteristics on the results. All cases use the model with AVRs at all generators and a single PSS connected to the generator at bus 1 from the previous subsection. 

Case~1 corresponds to optimization problem~\eqref{nlSDR} and the results presented in the previous subsection for the model with AVRs/PSS. In this case, both real and reactive power demand are decision variables but reactive power is tied to real power via a fixed power factor. Case~2 spatially shifts only real power demand (reactive power demand is fixed to the nominal reactive power demand), which achieves a slightly better optimal SDR. In contrast, Case~3 spatially shifts only reactive power demand (real power demand is fixed to the nominal real power demand), and we limit the change in reactive power demand at each bus with demand-responsive load to $\pm 100$ MVar. As shown in the table, optimizing the reactive power demand does not improve the SDR as much as optimizing the real power demand (24.1\% versus 36.5\% improvement).  Case~4 allows the real and reactive power demand to change independently; the SDR is greatly improved. Case~5 uses a smaller and more realistic limit on the change in reactive power demand ($\pm 20$ MVar at each bus with demand-responsive load) resulting in a lower optimal SDR than that of Case 4, but higher than that of Case~1, which does not allow real and reactive power demand to change independently. 

Case~6 allows only real power {\em generation} to change. The purpose of this case is to enable a comparison of SDR improvement due to demand changes to SDR improvement due to generation changes. As shown in the table, generator actions can achieve a larger percent improvement (41.2\% versus 24.1 -- 40.1\%). The greatest SDR improvement is achieved when we allow load and generation to change together (Case~7). In this case, the SDR improves by 49.5\%. However, in practice, generators may need more time than loads to respond, which can reduce the generators' ability to improve the SDR. For example, if we re-run Case 6 with ramp limits on the generators at buses 2-5 that restrict changes in real power generation to 1~MW, we obtain an SDR of 0.575\%, which is less than that achievable through DR, i.e., 0.696\%. Additionally, changing generator dispatch to improve the SDR will increase operational costs. 

\subsection{Comparison to Load Shedding}
We formulate and solve an optimization problem to determine the minimum load shedding needed to achieve the same SDR improvement as obtained in Case~1. Load shedding could be an alternate approach to improve small-signal stability, but it comes at a high cost and inconvenience to consumers.  The formulation of the minimum load shedding problem is 
\begin{linenomath}
\begin{subequations}
\begin{align}
\mathrm{min.}    \,\, &   \sum_{i\in \mathcal{S}_\mathrm{DR}} (p_{\mathrm{d},i}^0 - p_{\mathrm{d},i}) \\
\mathrm{s.t.}   \,\,  &  \eta_\mathrm{S} \geq 0.696\% \\
& \underline{p}_{\mathrm{d},i} \leq p_{\mathrm{d},i} \leq p^0_{\mathrm{d},i} \quad \forall i \in \mathcal{S}_\mathrm{DR}\\
& \mathrm{Constraints~}\eqref{nlSDR_e5}, \eqref{nlSDR_e7}-\eqref{nlSDR_e11}, \eqref{nlSDR_nl6}-\eqref{nlSDR_nl5}
\end{align}
\end{subequations}
\end{linenomath}
which we again solve using ILP with generalized eigenvalue sensitivities. We find that the system load (shown in the fourth row of Table~\ref{tab:14_pd}) would need to drop by at least 11\% to achieve the same SDR improvement as achieved by spatial load shifting. 

\subsection{Comparison to PSS Tuning}
We investigate whether it is  possible to improve the SDR simply through tuning the PSS gain, $K_{\mathrm{w},1}$. The maximum SDR achieved by tuning the gain (from 1 to 0.48) is 0.668\%, which is a 30.0\% improvement, smaller than that achievable through spatial load shifting. When we co-optimize the gain and the loading pattern, the SDR increases to 0.703\%, with $K_{\mathrm{w},1} = 1.12$ and the optimal loading pattern shown in the fifth row of Table \ref{tab:14_pd}. Note that the optimal loading pattern is similar to that obtained by spatial load shifting alone (third row of Table~\ref{tab:14_pd}) and the optimal gain is close to its previous value of 1. Tuning the gain {\em after} optimizing the loading pattern (i.e., fixing the loading to the values in the third row of Table~\ref{tab:14_pd} and then tuning the gain) results in an optimal $K_{\mathrm{w},1}$ of 1.01 and a SDR of 0.699\%. Therefore, in this case, once the loading pattern is optimized, tuning the PSS gain does not change the gain or improve the SDR significantly.

\subsection{Benchmarking Against a Simpler Method}
We compare the performance of our optimization-based spatial load shifting strategy with the simpler sensitivity-based generation redispatch strategy proposed in~\cite{chung2004generation}. To do this, we compute the sensitivity of the SDR to the real power generation numerically. Specifically, for each generator other than the generator at the slack bus, we determine the eigenvalues associated with the SDR at the current operating point $\lambda_m = \alpha_m \pm j\beta_m$ and also the eigenvalues associated with the SDR at a new operating point corresponding to a change in real power generation of generator $i$ by a small value $\Delta p_{\mathrm{g},i}$. We then compute the changes in the real and imaginary parts of these eigenvalues $\Delta \alpha$ and $\Delta \beta$, respectively.  Then, the sensitivity of the SDR to the real power generation of generator $i$ is 
\begin{linenomath}
\begin{equation}
\label{pg_sensitivity}
    SS_i = \frac{-\beta_m^2}{(\alpha_m^2+\beta_m^2)^{\frac{3}{2}}} \frac{\Delta \alpha}{\Delta p_{\mathrm{g},i}} + \frac{\alpha_m\beta_m}{(\alpha_m^2+\beta_m^2)^{\frac{3}{2}}} \frac{\Delta \beta}{\Delta p_{\mathrm{g},i}}.
\end{equation}
\end{linenomath}
Based on the sensitivities, we can change the outputs of the generators to achieve the desired improvement to the SDR.

Here, we find that the numerical sensitivities of the SDR to the real power generation of the generators at buses 2-5 are 0.0051\%, 0.031\%, 0.013\%, and 0.013\%, respectively. Since the generator at bus 3 has the largest sensitivity, we explore the impact of changing its real power generation on the SDR. Specifically, we attempt to achieve the SDR obtained by the optimization-based method through generation redispatch alone (Case 6), i.e., 0.726\%. This requires the generator at bus 3 to increase its real power generation by 6.8~MW according to its sensitivity. However, after implementing the change, the actual SDR achieved is only 0.697\%, which demonstrates the limitation of this simpler sensitivity-based approach.

\subsection{Understanding Optimal Loading Patterns}
\def\arraystretch{1.5}

\begin{table}[t]\footnotesize
    \centering
    \caption{Maximum SDR when shifting between two DR buses}
    \vspace{1em}
    \begin{adjustbox}{width=\columnwidth}
\begin{tabular}{|c||*{9}{c|}}
 \cline{1-2}
 3  & 0.6423 \\\cline{1-3}
5  & 0.5208 & \bf{0.6817} \\\cline{1-4}
6  & 0.5386 & 0.6005 & 0.5288 \\\cline{1-5}
9  & 0.5932 & 0.6054 & 0.5681 & 0.5278 \\ \cline{1-6}
10 & 0.5391 & 0.5958 & 0.5312 & 0.5191 & 0.5140 \\ \cline{1-7}
11 & 0.5228 & 0.6031 & 0.5199 & 0.5153 & 0.5157 & 0.5156   \\ \cline{1-8}
12 & 0.5290 & 0.5874 & 0.5238 & 0.5159 & 0.5152 & 0.5150 & 0.5140\\\cline{1-9}
13 & 0.5488 & 0.5917 & 0.5368 & 0.5188 & 0.5150 & 0.5148 & 0.5140 & 0.5141  \\\hline
14 & 0.5568 & 0.5667 & 0.5437 & 0.5237 & 0.5148 & 0.5146 & 0.5166 & 0.5165 & 0.5171 \\\hline\hline
 Bus \#   & 2 & 3 & 5 & 6 & 9 & 10 & 11 & 12 & 13  \\ \hline
\end{tabular}
 \end{adjustbox}
    \label{tab:SDR_twobus_1}
\end{table}

Lastly, in an effort to understand why loading patterns change the way they do when we maximize the SDR, we explore the impact of shifting load between each combination of two buses with demand-responsive load. To do this, we solve modified versions of \eqref{nlSDR} for the model with AVRs/PSS wherein each instance of the problem only allows shifting between two specific buses and no load shift limit is considered. The maximum SDR associated with shifting between each combination of two buses is shown in Table~\ref{tab:SDR_twobus_1}. The largest maximum SDR (bolded) is achieved when we shift load between buses 3 and 5. Two possible reasons for this are as follows. The first is that, by computing the participation factors \cite[p. 229]{Sauerbook} of the critical eigenvalue, we find that the critical eigenvalue is associated with the generator at bus 3 and so the SDR is particularly sensitive to the loading at bus 3. The second possible reason is that the combined load of buses 3 and 5 is the largest among all combinations.

\begin{table}[t]\footnotesize
    \centering
    \caption{Maximum SDR when shifting between two DR buses with equal loading ($p^0_\mathrm{d} = 15$~MW, $q^0_\mathrm{d} = 5$~MVar $\forall i \in \mathcal S_\mathrm{DR}$)}
    \vspace{1em}
    \begin{adjustbox}{width=\columnwidth}
\begin{tabular}{|c||*{9}{c|}}
 \cline{1-2}
 3  & 0.6928        \\\cline{1-3}
5  & 0.7009 & 0.6918        \\\cline{1-4}
6  & 0.7197 & 0.6934 & 0.6958       \\\cline{1-5}
9  & 0.7102 & 0.6978 & 0.7002 & 0.6937      \\ \cline{1-6}
10 & 0.7114 & 0.6975 & 0.6998 & 0.6938 & 0.6920      \\ \cline{1-7}
11 & 0.7147 & 0.6960 & 0.6983 & 0.6941 & 0.6932 & 0.6934     \\ \cline{1-8}
12 & \bf{0.7227} & 0.6947 & 0.6969 & 0.6922 & 0.6932 & 0.6933 & 0.6931       \\\cline{1-9}
13 & 0.7200 & 0.6950 & 0.6972 & 0.6928 & 0.6932 & 0.6934 & 0.6931 & 0.6918       \\\hline
14 & 0.7136 & 0.6971 & 0.6994 & 0.6942 & 0.6925 & 0.6919 & 0.6926 & 0.6935 & 0.6936\\ \hline\hline
 Bus \#   & 2 & 3 & 5 & 6 & 9 & 10 & 11 & 12 & 13  \\ \hline
\end{tabular}
 \end{adjustbox}
    \label{tab:SDR_twobus_2}
\end{table}

To eliminate the second reason and focus on understanding the first, we change the loading in the network to make the real and reactive power consumption at all buses with demand-responsive load equal, specifically,  $p^0_{\mathrm{d},i} = 15$~MW, $q^0_{\mathrm{d},i} = 5$~MVar $\forall i \in \mathcal S_\mathrm{DR}$, and we re-compute the SDR for each case. The results are given in Table \ref{tab:SDR_twobus_2}. The largest maximum SDR (bolded) is achieved when we shift load between buses 2 and 12 and the critical eigenvalue is associated with the generator at bus~2. We explored a number of nominal loading cases and found that, in most cases, the largest maximum SDR occurs when the bus connecting the generator associated with the critical eigenvalue is involved. This leads us to the hypothesis that  changing the load at this bus plays a key role in improving the SDR. Much remains to be done to fully understand the optimal loading patterns; we hope to pursue this in future work. A fuller understanding of these results could help us to develop heuristic spatial load shifting strategies to improve the SDR without needing to solve the optimization problem.

\section{Conclusion}\label{Sec: conclusion}
This paper developed a new DR strategy to improve small-signal stability. The optimal loading pattern was determined by solving an optimization problem using iterative linear programming and generalized eigenvalue sensitivities. We demonstrated the performance of the strategy via case studies using the IEEE 14-bus system. We showed that the SDR is improved by spatially shifting load and the improvement is greater than that achieved by slow-acting generators. We also found that use of different system models resulted in different optimal loading patterns and that neglecting AVR and PSS models can cause a significant reduction in the SDR, suggesting the importance of properly modeling the system in the formulation. Additionally, we benchmarked our approach against to a simple sensitivity-based generation redispatch method, and showed that our optimization-based approach can achieve a better stability improvement. 

Future work will seek to gain a better understanding of the optimal loading patterns. Our preliminary investigation pointed to the importance of the generator associated with the critical eigenvalue, but more work is needed to understand optimal loading patterns and develop computationally-simpler heuristic spatial load shifting strategies. In the future, we will also consider including different types of load models and also N-1 security constraints in the formulation. Other avenues for future work include combining multiple stability objectives into a single optimization problem to avoid circumstances in which improving one type of stability negatively impacts another. We are also interested in developing DR strategies to improve transient stability.

\section*{Acknowledgment}
The authors thank Prof.~Ian Hiskens for helpful suggestions, Kasra Koorehdavoudi for initial collaboration on this topic, and Muyang Liu for suggestions on computing eigenvalues.

\bibliographystyle{model1-num-names}
\bibliography{Ref}

\end{document}